\documentclass{aa}
\usepackage{epsf,rotate}

\begin{document}

   \thesaurus{ 06
              (08.02.4;           % binaries: spectroscopic
               08.09.2 V1500 Cyg; % Stars: individual
               08.14.2;           % novae, cataclysmic variables
               08.02.1            % (stars:) binaries(including multiple): close
               08.02.2;           % binaries: eclipsing
               08.23.1)           % white dwarfs
   }
\title{Post nova white dwarf cooling in V1500 Cygni}

   \subtitle{}
   
   \author{M.W. Somers \and T. Naylor}
                
   \offprints{T. Naylor}
                   
   \institute{Dept. of Physics, Keele University, Keele, Staffordshire, ST5 5BG}

   \date{Received ???; accepted ???}

   \maketitle

\begin{abstract}
We use the irradiated secondary star in the remnant of Nova
Cyg 1975 as a probe of the cooling white dwarf.
At superior conjunction the flux in the B band is dominated by the irradiated
face of the secondary star.  
The heated face produces an orbital modulation which is dependent on the 
strength of the irradiating source.
We demonstrate that the cooling rate of the white dwarf is consistent with the
theoretical model of Prialnik (1986).
\end{abstract}


\section{Introduction}
\label{intro}
In many close binary systems the hotter star will heat one face of the cooler
star.
As the orbital motion of the binary brings this face into view the observed
flux from the binary will increase, only to fall again as it rotates out of
view.
After a nova explosion, the hot white dwarf is an obvious candidate for 
heating its cool companion.
Probably the best evidence is in V1500 Cyg (Nova Cyg 1975), where Schmidt 
et al. (1995) show that the secondary star dominates the photometric 
modulation.
They fitted HST spectra with a red star whose unperturbed temperature was 
$\simeq 3000K$ but whose face towards the white dwarf was $\simeq 8000K$.

There are three other novae for which there is photometric evidence for
heating.
The orbital modulation in DN Gem (Nova Gem 1912) was found by Retter,
Leibowitz \& Naylor (1999) to be well described by a heating model and
DeYoung \& Schmidt (1994) suggested heating could explain the lightcurve of 
V1974 Cyg (Nova Cyg 1992).
Finally, Somers, Mukai \& Naylor (1996) found the infrared lightcurve of WY 
Sge (Nova Sge 1783) required a heated face to be modelled successfully.
However, in this case the level of irradiation was so low that it could
have been supplied by the accretion luminosity, as occurs in the dwarf nova
IP Peg during outburst (Webb et al 1999).

This led us to ask if there were further evidence available which would help
us identify the source of irradiation in old novae as the white dwarf.
It is particularly important to do so, as the irradiated surface of the
secondary star may be the most reliable diagnostic we have of the 
white dwarf luminosity, since its intrinsic radiation is produced in the far 
UV/soft X-ray regimes, where the effect of interstellar absorption will
be very marked.  
If the irradiating object is the white dwarf, the irradiation should
decrease on the white dwarf cooling timescale. 
Prialnik (1986) shows how the surface layers of a white dwarf are heated
during a nova explosion and cool as a power law on a time scale of 200 yrs.  
(This is in
contrast to the cooling of white dwarfs after their initial formation, which
involves cooling of the entire star and occurs on a time scale of $10^{8}$
years.)  

\section{History of the photometric modulation in V1500 Cyg}
\label{5:history}

V1500 Cyg was a naked eye nova, reaching a peak magnitude of V=2.2 in late
August 1975.
It is thought to be currently a slightly asynchronous AM Her system.
The magnetic nature of the white dwarf was first proved by the
detection of circularly polarized light (Stockman et al. 1988).  
The present day photometric period (orbital) is $1.8\%$ longer than the
polarimetric period (white dwarf spin).  
Presumably the two were thrown out of equality by the nova explosion and 
observations suggest that they will re-synchronize in a time scale of ~200 
years (Schmidt \& Stockman 1991).

Stockman et al. (1988) explain the photometric period  evolution as follows:

$\bullet$ The expansion due to the nova explosion increases the moment of
inertia of the white dwarf, causing its spin period to increase to 0.141d,
breaking the synchronous rotation.
At this point the photometric modulation is associated with the spin
period of the white dwarf.

$\bullet$ Interaction between the secondary star and the envelope causes the 
envelope to be spun up to the binary period, strong coupling ensures that the 
core also achieves synchronism.

$\bullet$ The remnant envelope shrinks back onto the white dwarf surface, 
reducing its moment of inertia and thus decreasing the spin period to 0.137d.

\vspace{1cm}
Currently, therefore, V1500 Cyg displays two periods.
There is a polarimetric signal at 0.137d which is the spin period of the
white dwarf, and a photometric signal at 0.140d which is the orbital
period of the binary.
Aside from Schmidt et al's spectrophotometry outlined in Section \ref{intro},
further evidence that the secondary star now dominates the photometric 
modulation comes from the fact that  the timing of flux maximum in our own 
data matches the orbital ephemeris (see Section \ref{5:obs}).
It is also clear however, from the presence of flickering and the slightly 
asymmetrical B band light-curves, that there is some contamination of the 
orbital modulation.
This could either be due to the accretion stream, or perhaps because the
spin period of the white dwarf may also have a photometric signal
(Pavlenko \& Pelt 1988).

\section{Modelling white dwarf cooling after surface heating.}
\label{5:modelling}
Prialnik (1986) modelled the evolution of a classical nova through a complete
cycle; accretion, outburst, mass loss, decline and resumed accretion.  The
model is for a 1.25$M_{\odot}$ C-O white dwarf.  
The resulting outburst is fast and similar to that observed in V1500 Cyg, 
matching the composition of the ejected envelope very well.  
The white dwarf is modelled allowing for heat transfer via radiation, 
conduction (Iben 1975, Kovetz \& Shaviv 1973) and convection (Mihalas 1978).
The result may be fitted well with a power law cooling curve of the form
\begin{equation}
L\propto t^{-1.14},
\end{equation}
where L is the luminosity of the white dwarf and $t$ is the time since
outburst.

\section{Observations}
\label{5:obs}                                        

To extend the baseline of photometric amplitude decline we obtained
one orbital cycle of B and V band photometry using the JKT on La Palma on the
night of 1995 October 3.   
The TEK4 CCD was used with pixels binned 2 by 2 to achieve 0.6"x0.6" pixels 
in rapid readout mode.   
The seeing was around 2.0 arcsec.
The exposures were typically of 120 seconds with filters being alternated
between observations.

\begin{figure}
\rotate[l]{\epsfxsize=60mm
\epsffile{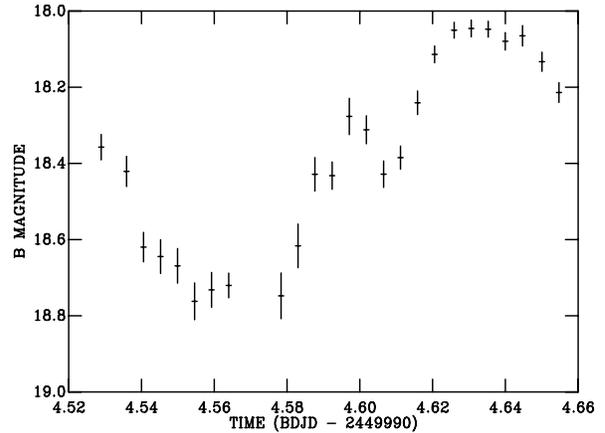} }
\caption{The B band magnitude of V1500 Cyg plotted against Barycentric 
Dynamical Julian Date.}
\label{fig:bband}
\end{figure}

\begin{figure}
\rotate[l]{
\epsfxsize=60mm
\epsffile{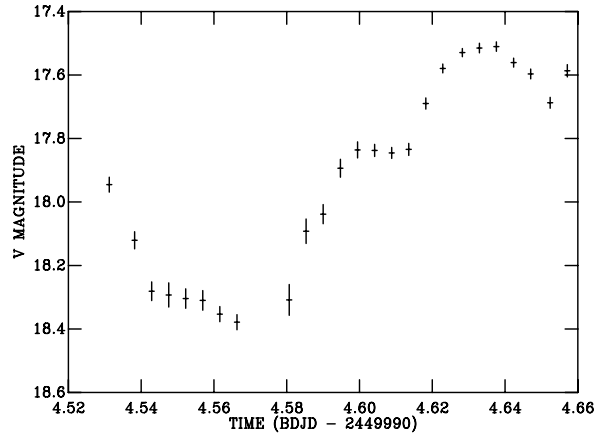} }
\caption{The V band magnitude of V1500 Cyg plotted against Barycentric 
Dynamical Julian Date.}
\label{fig:vband}
\end{figure}

A bias frame was subtracted from each image, and it was then flatfielded using 
an image of the twilight sky.
The counts for V1500 Cyg and various other stars were extracted from each
frame using the optimal weighting procedure described in Naylor (1998).
We divided the counts for V1500 Cyg by those for star C1 of Kaluzny
\& Semeniuk (1987), allowing us to put the lightcurves in Figures 
\ref{fig:bband} and \ref{fig:vband} onto a magnitude scale.

The times of maximum and minimum agree, within errors, with the ephemeris 
of Semeniuk, Olech \& Nalezyty (1995).
The data do not show a pure heating modulation since there is some evidence
of a dip in the light curve at time 2\,449\,994.61.
Such dips are not uncommon, see for example the light-curves of Kaluzny \&
Semeniuk (1987) whose data are from 10 years earlier than our observations.
These irregularities add  errors in the estimation of the amplitude of the
modulation (see Section \ref{5:Ampvtime}).

\section{The Amplitude v Time relationship} \label{5:Ampvtime}

\begin{figure}
\rotate[l]{
\epsfxsize=60mm
\epsffile{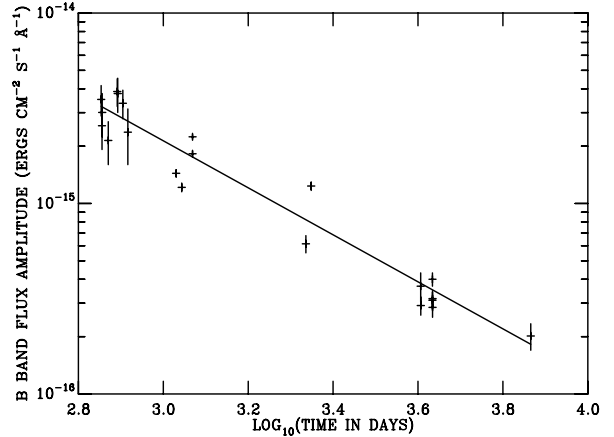} }
\caption{The decaying photometric variability of V1500 Cyg}
\label{fig:ampVd}
\end{figure}

By searching through various sources a record of the B band photometric 
behaviour observed in V1500 Cyg since outburst has been assembled.  
Table 3.1 gives a list of all references used.
The amplitude of the B band photometric modulation versus the time since
outburst is plotted in Figure \ref{fig:ampVd} on logarithmic scales.
We immediately note that these points lie roughly in a straight
line, implying a power law decay in time, although it should be noted 
that the first year of data have been omitted.
While other authors have attempted to find a trend in the amplitude versus
time data they relied on the amplitude in magnitude space.  
This is a measure of the amplitude relative to the brightness of other parts 
of the system.  
Clearly this is inappropriate for V1500 Cyg since the overall brightness in 
the B band changes due to a large variety of effects.
The amplitude in flux space is a good probe of the irradiation 
since it does not need to be corrected for changes in
relative brightness of other parts of the system.  
Hence, what we refer to as the ``flux amplitude'' is obtained by converting
the magnitudes at orbital maximum and minimum into fluxes (assuming a
star of $B$=0.0 gives $7.2 \times 10^{-9}$ ergs cm$^{-2}$ s$^{-1}$ \AA$^{-1}$), and differencing
them.
An unweighted straight line to these data give
\begin{equation}
\label{eqt:observe}
A \propto t^{-1.26\pm0.21},
\end{equation}
where $A$ is the flux amplitude, and the error bar corresponds to 1$\sigma$.

\begin{table} \label{table:v1500refs}
\caption{The flux in the orbital modulation of V1500 Cyg.}
\begin{center}
\begin{tabular}{lcc}
Days from  &  Flux amplitude   & \\
outburst   &  (10$^{-16}$ ergs cm$^{-2}$ s$^{-1}$ \AA$^{-1}$)  & Reference\\
\\
320  &$ 1060 \pm 200 $& Kemp, Sykes \&                \\
329  &$  720 \pm 190 $&    Rudy (1977)                \\
357  &$  310 \pm 32 $& Patterson (1978)              \\
714  &$   35 \pm 6 $&      {\tt "}                  \\
717  &$   26 \pm 6 $&      {\tt "}                  \\
718  &$   30 \pm 8 $&      {\tt "}                  \\
742  &$   21 \pm 5 $&      {\tt "}                  \\
779  &$   39 \pm 6 $&      {\tt "}                  \\
781  &$   38 \pm 8 $&      {\tt "}                  \\
803  &$   34 \pm 6 $&      {\tt "}                  \\
825  &$   24 \pm 8 $&      {\tt "}                  \\
1072 &$   14 \pm 0.6 $& Patterson (1979)              \\
1105 &$   12 \pm 0.6 $&      {\tt "}                  \\
1172 &$   22 \pm 0.6 $&      {\tt "}                  \\
1172 &$   18 \pm 0.6 $&      {\tt "}                  \\
2167 &$   61 \pm 0.6 $& Horne \&                      \\
2230 &$   12 \pm 0.6 $&   Schneider (1989) $^\dag$    \\
4039 &$    3.6 \pm 0.6 $& Kaluzny \&                    \\
4041 &$    2.9 \pm 0.3 $&     Semeniuk (1987)           \\
4305 &$    3.1 \pm 0.3 $& Kaluzny \&                    \\
4306 &$    3.9 \pm 0.3 $&    Chlebowski (1988)          \\
4307 &$    2.8 \pm 0.3 $&      {\tt "}                  \\
4308 &$    3.1 \pm 0.3 $&      {\tt "}                  \\
7339 &$    2.0 \pm 0.3 $& This Work                     \\
 &$ $& \\

\end{tabular}
\end{center}
$^\dag$These points are from spectroscopy, band pass 4750-4790\AA.
\end{table}

It is apparent that there is an extra source of noise that acts on a short
time scale compared with the general trend of amplitude decline. 
(Hence our choice of an unweighted fit.) 
This is likely to be due to the aforementioned orbital dips shifting around 
the orbital light-curve.
The position of maximum light from the accretion columns with respect to
the secondary star's photometric hump will vary with beat phase.
This will cause the apparently
random excess scatter in the photometric amplitudes.

\section{The Irradiating Flux v Time relationship} \label{5:Firrvtime}
The bolometric flux amplitude from the secondary star will be proportional 
to the irradiating flux.
However, the flux measured in any particular bandpass will not follow such a
proportionality, as the bolometric correction will change with the temperature
of the heated face.
To extract how the irradiation is changing from the changing flux
amplitude one must determine the response of the heated face of the
secondary star to decreasing the heating.  One may conveniently approximate
the response as 
\begin{equation}
\label{eqt:response}
A \propto F_{irr}^{x} \propto t^{x  \eta}.
\end{equation}

Here $\eta$ is the power law decline predicted by Prialnik to be -1.14 while
$x$ is the response of the secondary star to heating.   
In fact $x$ is not a constant but is a weak function of the temperature of 
the heated face of the secondary star.
We therefore used the code described in Somers et al (1996) determine how the
amplitude of the modulation changes with the irradiating flux.
This model irradiates a Roche-lobe filling star from a point source at
the position of the white dwarf.
The irradiation then raises the temperature of each surface element in
impinges on, such that all the incident energy is re-radiated.
We will also assume that the radiation from the secondary star can be
approximated as a blackbody, an assumption we shall examine later.
We used a mass ratio q=3, inclination i=$60^{\circ}$ and underlying
secondary star temperature of 3000K.  
These values are typical of those quoted for V1500 Cyg and the results are, 
in any case, insensitive to the exact values.  

We began by irradiating a 3000K secondary star such that the flux when the
irradiated face was towards the observer was equivalent to a star of
8000K, matching the front face temperature observed by HST.
This model should correspond to the last data point in Figure \ref{fig:ampVd}.
The flux amplitude in Figure \ref{fig:ampVd} declines by a factor of 10$^{1.2}$,
and so we increased the irradiation in our model until the flux amplitude
had increased by this factor.
Over this range of interest, we found that 
\begin{equation}
A \propto F_{irr}^{0.75} 
\end{equation}
represented the data to better than 25 percent for all values of $F_{irr}$.
We also used tried using the bolometric correction and colours of model 
atmospheres given in Bessell et al (1998) instead of blackbodies to
represent the flux.
We found this changed $x$ by less than 0.06 in the low irradiation case,
which is that most affected by the difference between model atmospheres
and black bodies.

With a value for $x$ we can now use the observations to derive a value of 
$\eta$ by equating (\ref{eqt:observe}) to (\ref{eqt:response}).
This yields:
\begin{equation}
\eta=0.94\pm0.09.
\end{equation}
This result is just (2.2$\sigma$) consistent with the value Prialnik (1986).
find from purely theoretical considerations of $\eta=1.14$.
Especially given the nature of the approximations made, this seems to support
the conclusion that the photometric variation in V1500 Cyg, and by
implication other old novae, is caused by irradiation from the white dwarf.

\section{Conclusions} \label{5:conclusions}
                 
The above shows that for at least the first 20 years from outburst the
white dwarf cooling models match the available observations.
Given the Prialnik-type cooling law, with the observed value of $\eta$,
and the temperature of the irradiated face at some known time after outburst
(from Schmidt et al. 1995) then we can calculate the typical time taken for
irradiation of the surface to become negligible.
The irradiation will drop off so that the incoming radiation is less than 
double the unheated surface luminosity of the secondary star about $280\pm140$
years after the outburst of V1500 Cyg.  
Interestingly, we find that in WY Sge, now over 200 years since nova outburst,
the irradiation from the white dwarf has declined to these levels
(Somers et al 1996), although in that system the disc is a complicating factor.
Thus both V1500 Cyg, and WY Sge suggest that white dwarfs really do cool as
the theory predicts.

\section*{ACKNOWLEDGMENTS}

The Jacobus Kapteyn Telescope is operated on the island of La Palma by the 
Isaac Newton Group in the Spanish Observatorio del Roque de los Muchachos 
of the Instituto de Astrofisica de Canarias.
We thank Gregory Beekman and Coel Hellier who helped with the observations, 
and Alon Retter for commenting on the manuscript.
TN was in receipt of a PPARC advanced fellowship when the majority of this
work was carried out.

\end{document}